\documentclass[conference]{IEEEtran}
\IEEEoverridecommandlockouts
\usepackage{cite}
\usepackage{amsmath,amssymb,amsfonts}
\usepackage{algorithmic}
\usepackage{graphicx}
\usepackage{textcomp}
\usepackage{xcolor}
\usepackage{listings}
\usepackage{url}
\usepackage{flushend}
\def\BibTeX{{\rm B\kern-.05em{\sc i\kern-.025em b}\kern-.08em
    T\kern-.1667em\lower.7ex\hbox{E}\kern-.125emX}}
\begin{document}

\title{Evaluating an Effective Ransomware Infection Vector in Low Earth Orbit Satellites\\

\thanks{This work was funded by the Department of Computer Science, MTU Cork.}
}

\author{\IEEEauthorblockN{Marin Donchev\IEEEauthorrefmark{1},
Dylan Smyth\IEEEauthorrefmark{2}}
\IEEEauthorblockA{\textit{Department of Computer Science} \\
\textit{Munster Technological University}\\
Cork, Ireland \\
Email: \IEEEauthorrefmark{1}marin.donchev@mycit.ie,
\IEEEauthorrefmark{2}dylan.smyth@mtu.ie}}

\maketitle

\begin{abstract}
Non-Terrestrial Networks (NTNs) and satellite systems have become an important component of modern data communication systems in recent years. Despite their importance, the security of these systems is often limited, leaving them vulnerable to determined attackers. In this paper, we outline a scenario in which an attacker can infect an in-orbit NASA Core Flight System (cFS) based satellite with ransomware and communicate the infection back to a satellite operator. This paper is the first to demonstrate an end-to-end exploit path that results in a ransomware infection without the need for a supply chain attack or compromised credentials. Novel ransomware is delivered to an emulated satellite system using custom shellcode that exploits a weakness in the considered scenario. The scenario considered by this initial piece of work achieves a success rate of 33.3\% for a complete successful infection. 
\end{abstract}

\begin{IEEEkeywords}
Non-Terrestrial Networks, Satellite, Security, Malware, Ransomware
\end{IEEEkeywords}

\section{Introduction}
The number of Low Earth Orbit (LEO) satellites deployed for civil, commercial, government, and military use has increased significantly over the past decade\cite{al2022survey, ucssatellite}, with 7,701 being in operation as of  2024 \cite{oribitingnow}. This is in part a result of increased use of Non-Terrestrial Networks (NTNs) for internet communication, but also due to lower costs in building and deployment\cite{davoli2019small, adilov2022analysis}. Satellites deployed for specific data collection purposes previously relied on specialized embedded processors for data and communications processing. ARM-based systems, running Real-Time Operating Systems (RTOS), are now popular due to their high processing and general computing capabilities. While the use of ARM and RTOS in satellites has introduced many opportunities for the mixed use of satellite deployments, it also presents challenges for security. 

The security of satellite infrastructure has largely relied on security through obscurity, where specialized knowledge and equipment was needed to interact with satellite systems. However, this knowledge has become more open over previous years, and a growing interest in Software-Defined Radio (SDR) equipment has led to an increase in affordable, consumer focuses devices. Accessibility of knowledge and equipment, paired with the the popularity of LEO satellite systems, who's proximity to the ground make connectivity from a bad actor possible, increases the risk that these systems will be targeted. A threat that poses a significant risk to these systems is a ransomware attack, where an attacker will disrupt the operations of a system until a ransom has been paid. Examples of satellite ransomware have been presented in literature, but this work only considers post-exploitation scenarios. No work has previously examined the technical circumstances required for an attacker to infect a satellite and communicate this infection to a satellite operator.

This paper examines how an attacker can compromise an in-orbit satellite, infect that satellite with effective ransomware, and communicate the infection to the system operator. This work focuses on how a security vulnerability can be introduced by a programmer when building a custom application for NASA's Core Flight System (cFS)\cite{mccomas2016core}. Guided by the SPARTA framework \cite{SPARTA2024}, scenarios based around the Common Weakness Enumeration (CWE) Top Ten Known Exploited Vulnerabilities list \cite{CWETop102023} are considered and one of these scenarios is selected. Novel ransomware, which takes control of the radio module used in our testbed, is then developed, along with an exploit to deliver and execute this ransomware. Methods to prevent this attack, even when a vulnerability is present, are explored.

The remainder of the paper is structured as follows. Section \ref{sec:related_work} presents a literature review of satellite system security. Section \ref{sec:attack_scenario} provides details on our considered attack scenario, ransomware, and exploit. Section \ref{sec:evaluation} presents an evaluation of the success rate of the considered attack scenario. Section \ref{sec:discussion} discusses the work, and finally a conclusion is presented in Section \ref{sec:conclusion}.

\section{Related Work}
\label{sec:related_work}

\subsection{Satellite Attack Vectors}
The components of a NTN and satellite system can be defined as the \textit{Ground Component} and \textit{Space Component}, with the \textit{Communication Channel} transferring data between them. There are real-world examples of attacks targeting each of these components \cite{AsatCyber, gurantz2024satellites, HackerBroadcast, ViasatHack}. These real-world examples largely involve jamming or unauthorized communication with satellite system components.

Limited work has been done in the area of identifying potential attacks against the space component of satellite systems. This is due to a significant amount of the technology being close-sourced and held behind trade secrets \cite{pavur2022building}. Pavur \textit{et al.}\cite{pavur2022building} document the threat of signal jamming and it's role in Denial of Service (DoS) attacks. Willbold \textit{et al.} \cite{willbold2023space} focus on software vulnerabilities and present results from reverse engineering three satellite firmware images, used in real spacecraft. The authors found numerous buffer overflow vulnerabilities, some of which were exploited for demonstration. 

As there is a difficulty in accurately identifying attack vectors on real systems, work has been done to develop security frameworks which identify potential attack vectors and security considerations\cite{SPARTA2024}. However, it has been noted that items in this framework lack validation \cite{curbo2023research}.

Work in the area of attacks against the space component has also been limited by the difficulty in accessing satellite systems, both by security researchers and potential attackers. This, however, is changing with the increase in popularity and power of Software-Defined Radios (SDRs). Lin \textit{et al.} \cite{lin2022defending} highlight that attackers are able to use SDRs to eavesdrop packets sent to a satellite and replay or edit them. Work has also shown that a functional GS can be developed with publicly available equipment for a low cost \cite{willbold2023space ,singh2021community, HackSatVideo}. Greater accessibility to satellite systems demonstrates the need to ensure that those systems and secure and resiliency against attacks.

\subsection{Satellite Malware}
Falco \textit{et al.} \cite{falco2023wannafly} present a concept of a ransomware for NASA's cFS by flooding the Software Bus (SB) with messages, rendering it unusable for internal cFS applications and disrupting the spacecraft's mission. Hansen, Falco \textit{et al.} \cite{hansen2024guarding} continue this work by implementing a malicious version of the File Manager application for cFS to flood the SB with messages until a special key is inputted. While these works demonstrate the potential of malware that has infected a cFS device, the malware developed by the authors must be compiled along-side cFS, restricting the infection vector to a supply-chain attack. 

While work has been done in identifying security issues in each component of a satellite system, no work has previously considered how an attacker can leverage a vulnerability, such as those described in \cite{willbold2023space}, to infect a satellite with ransomware and how that ransomware can achieve it's goal without having been introduced to the system in the supply-chain.
\section{Attack Scenario}
\label{sec:attack_scenario}
In this section the threat model, considered scenario, and exploit path are detailed.

% What assumptions do we make about the attacker (e.g. they have access to a radio that can transmit to the satellite) 
\subsection{Threat Model}
This work considers an attacker who has the capability, through access to a custom rogue GS, to communicate with a target satellite. The attacker has knowledge of the target satellites software, including the vulnerability present in that software, and they are familiar with the methodology used to exploit the vulnerability. The goal of the attacker is to infect the satellite with ransomware designed to target the system.

% Testbed architecture. Note the use of cFS, radios, and the ground station software here.
\subsection{Satellite System}
\label{sec:satellite_system}
\begin{figure}[!hbt]
    \centering
    \includegraphics[width=1.0\linewidth]{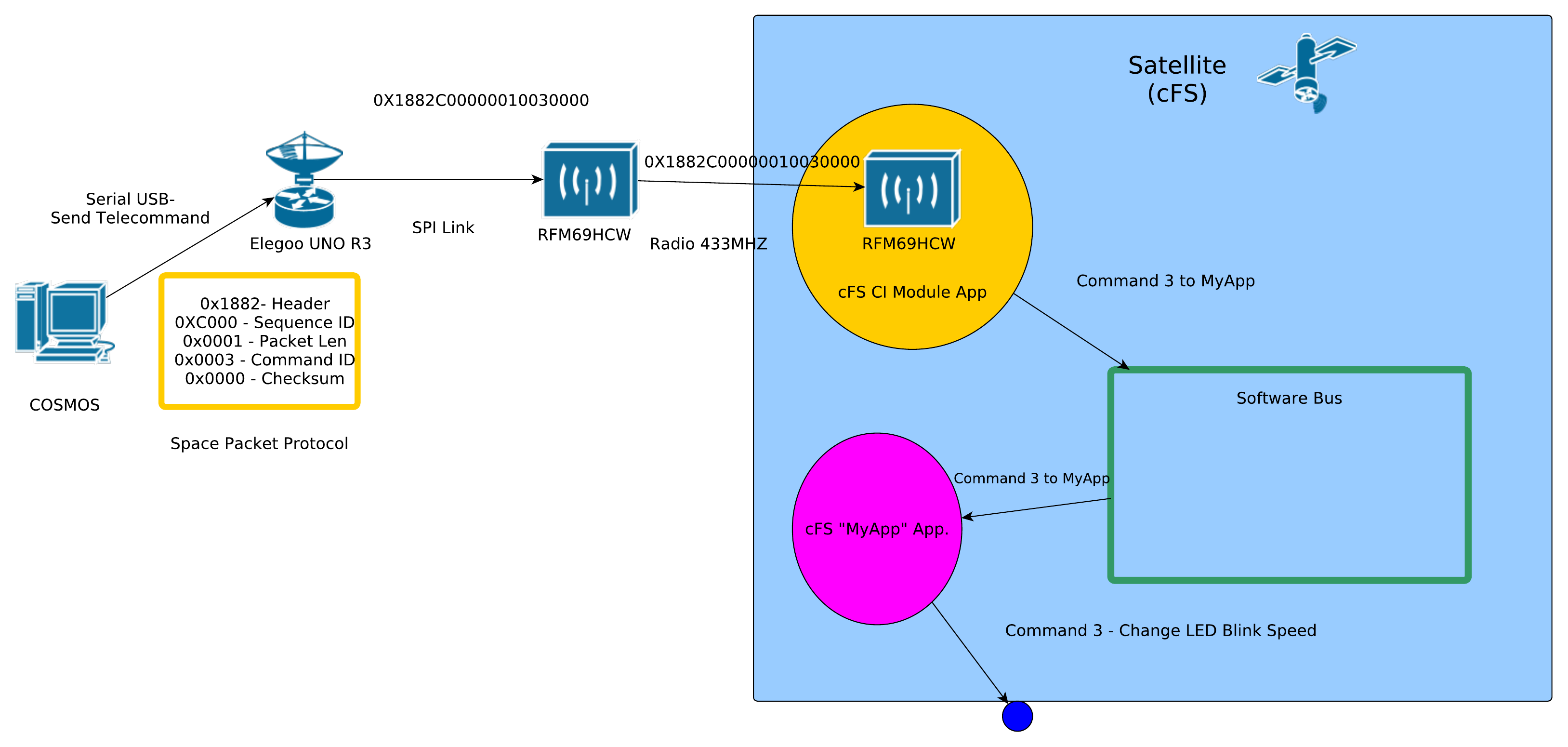}
    \caption{Communication between the components of the testing environment}
    \label{fig:testing-env-comms}
\end{figure}

The architecture of the scenario considered in this paper is provided in Figure \ref{fig:testing-env-comms}. The satellite used in this scenario is running NASA's cFS with a custom application designed to react to commands from the GS component. As the satellite is running cFS, it is considered to be built using a Single Board Computer (SBC) running a Linux-based operating system. To enable rapid prototyping of the ransomware used in this scenario, we also consider that Python is available on the Satellite's operating system. The availability of Python on the system and it's role in this scenario is discussed in Section \ref{sec:discussion}. OpenC3's COSMOS \cite{openc3COSMOSNASA} is used as the GS software. The satellite communicates with the GS using a radio module with no encryption present. The attacker communicates with the satellite using a similar radio module integrated into a rogue GS.

To implement this system design as a testbed, a Raspberry Pi 4B was used to emulate the satellite, running a 32-bit Raspbian and cFS. The default CI module code was edited to utilize an RFM69HCW radio, communicating at 433MHz and connected via SPI. The GS was emulated using a Linux host running OpenC3's COSMOS, configured to use CCSDS Space Packet Protocol. The radio is connected to the emulated GS using an Elegoo Uno R3 board as a bridge. The attacker's rogue GS is emulated using the same configuration.

% \subsubsection{Satellite}
%The flight software, chosen for the research and development of this work, is NASA's cFS. It is a popular choice for research of cybersecurity for space, as demonstrated by the works of Schalk et al. \cite{schalk2023detection} and Falco, Hansen et al. \cite{falco2023wannafly}, \cite{hansen2024guarding}. There are also other choices such as "CubeSatSim" \cite{githubGitHubAlanbjohnstonCubeSatSim} but cFS stands out as a more ubiquitous choice. The version used is 6.7.0, the "Aquila Bundle", with the default applications that come with it. The CI module application has been modified to use radio communication rather than UDP. 

% \subsubsection{Ground Station}
% The GS software for the testing environment is OpenC3's COSMOS, version (version 5.14.0). The distribution and it's integration with cFS can be viewed at \cite{openc3COSMOSNASA}. Modifications have been made to the cFS target so that USB serial connection is used in order to communicate with a radio module.

% Vulnerabilities:
% 1. No encryption for communication
% 2. Programming error that leads to a cFS app being vulnerable to a buffer overflow.

\subsection{Vulnerability Emulation}
\label{vuln-emu}
The vulnerability considered in this scenario exists in the custom application included in cFS on the satellite. The Command Ingest (CI) path in the custom application is designed to buffer data send from the GS. We consider that a vulnerability has been introduced here through the use of unsafe C functions. The line, shown in Listing \ref{lst:vulnerability}, is used to unsafely copy the data ingested from the GS communication into a new buffer, leading to a potential buffer overflow vulnerability.

\begin{lstlisting}[
caption={Buffer Overflow Vulnerablity},
label={lst:vulnerability}
]
memcpy(data_buf, ingest_buf, len_to_copy);

\end{lstlisting}

Here, the \textit{ingest\_buf} variable is copied directly into \textit{data\_buf} without checking the length first. To simplify experimentation, the \textit{data\_buf} buffer is only 10 bytes long and can be easily overflown. This allows the attacker to send a payload, using the radio, which can overflow the buffer and overwrite the return address and allow for an Remote Code Execution (RCE).
% For this concepts, we assume that a programming error has occurred during the development of the CI module of cFS, in a similar manner as described by Willbold et al. \cite{willbold2023space}. Since cFS is written in C, a buffer overflow can be introduced. The particular mistake is in the following line: 

% \verb|memcpy(vuln, buf, len_to_copy);|

% Here, the \textit{buf} variable is copied directly into \textit{vuln} without checking the length first. The \textit{vuln} buffer is only 10 bytes long and can be easily overflown. This allows the attacker to send a payload, using the radio, which is big enough to exploit this and overwrite the return address and allow for an RCE. For this scenario, we also assume that there is no encryption in the communication channel.

% How the malware works and how it's different to other satellite malware designs described in the related work.

\subsection{Ransomware Design}
\begin{figure}[!hbt]
    \centering
    \includegraphics[width=1.0\linewidth]{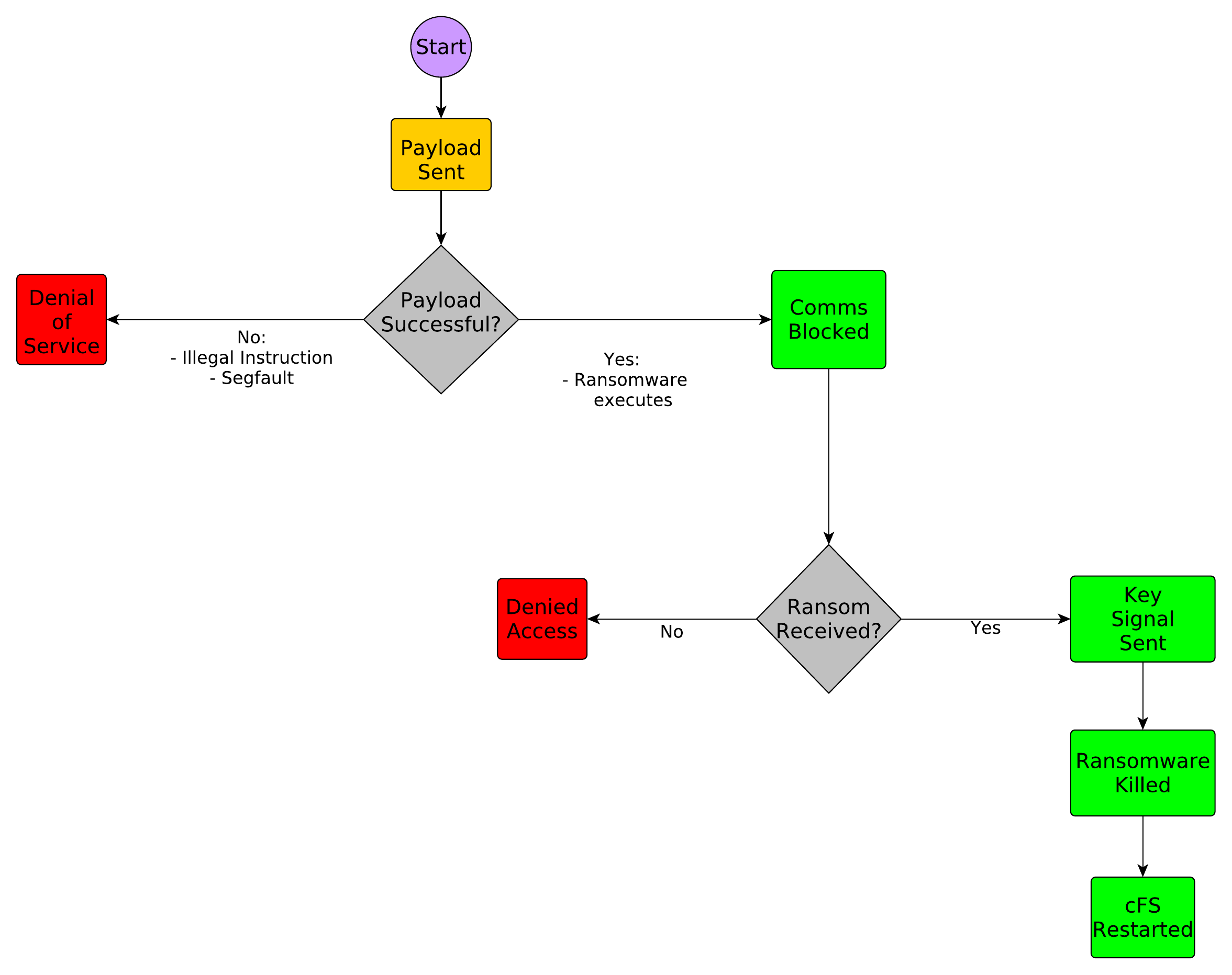}
    \caption{Malware execution flow}
    \label{fig:malware-flow}
\end{figure}

The execution flow for the ransomware designed for this work can be seen in Figure \ref{fig:malware-flow}. The ransomware operates by reinitializing and taking control of the radio used for the satellite-to-GS communication. By taking control of this component, the critical components of cFS can continue to operate while messages sent from the GS to these components are blocked. The ransomware uses the radio to notify the operator at the GS of the infection while also listening for an expected secret key used to disable the ransomware. Once the correct key, available to the operator after ransom payment, is received, the ransomware will restart cFS to resume normal satellite operation and then terminate itself.

% The execution flow can be seen on Figure \ref{fig:malware-flow}. Once the payload executes successfully, it will re-initialize the radio under the ransomware's control. The ransomware will ignore any messages received except a special key which only the attacker possesses. It uses the radio to notify the operators of the ongoing ransom and where to pay. Once payment is received, the attacker can send the key message to the satellite using radio. The ransomware will restart cFS to resume normal satellite operation and then terminate

% \begin{figure}[!hbt]
%     \centering
%     \includegraphics[width=1.0\linewidth]{figures/malware-flow.pdf}
%     \caption{Malware execution flow}
%     \label{fig:malware-flow}
% \end{figure}

% 1. How do you exploit the buffer overflow,
% 2. How is the malware delivered.
% 3. How is the malware executed.

\subsection{Exploitation}
Exploitation of the vulnerability described in \ref{vuln-emu} is done using custom shellcode. Communication with the satellite is performed using a rogue GS. The payload is delivered across multiple messages and buffered at the CI of the custom cFS app. Once copied into the vulnerable buffer, padding and a NOP sled is used to correctly align the shellcode to the instruction pointer. The Python-based ransomware is directly embedded within the shellcode. Upon execution, the shellcode performs the following actions:
\begin{enumerate}
    \item Open system call to create the an empty Python file on the satellite's file system and retrieve a file descriptor.
    \item Write system call to write the embedded ransomware code to the newly created file.
    \item Close system call on the new file.
    \item Use the execve system call to execute the ransomware file on the satellite.
\end{enumerate}

One executed, the ransomware takes control of the satellite's radio and consumes all future messages until the correct key is sent from the COSMOS GS.

\section{Evaluation}
\label{sec:evaluation}
The goal of the evaluation for this work was to confirm that the attack and ransomware operated as expected, and to evaluate the reliability of the exploit. 

\subsection{Methodology}
To evaluate the effectiveness of the attack, the cFS app, described in Section \ref{vuln-emu}, is installed on the emulated satellite. The exploit is delivered to system using the rogue GS, and the output of the cFS app, the satellite's file system, and the COSMOS GS are observed to establish successful exploitation of the vulnerability, and the successful delivery and execution of the ransomware. To evaluate the reliability of the exploit, the above process is repeated 300 times to identify a success and failure rate for the entire attack.

\subsection{Results}
\begin{figure}[!hbt]
    \centering
    \includegraphics[width=0.9\linewidth]{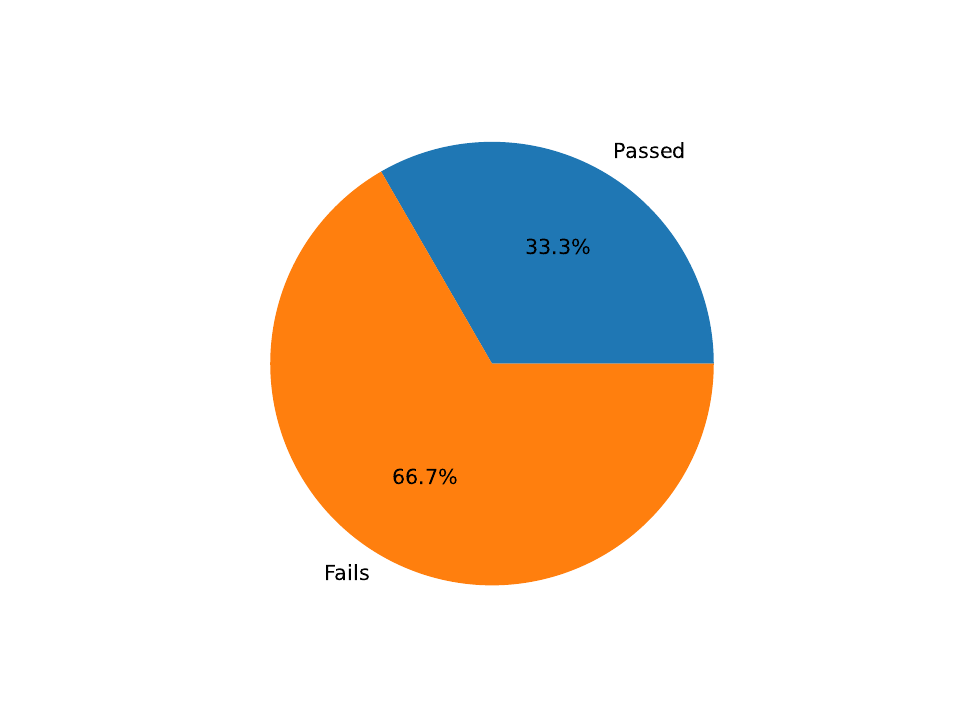}
    \caption{Success rate of the exploit}
    \label{fig:pie-success-fail}
\end{figure}

In every test where exploitation was successful, the ransomware was found to be effective in taking control of the GS-to-satellite communication. The success rate of the exploit is shown in Figure \ref{fig:pie-success-fail}. While the success of the post-exploitation steps of the attack was affirmed, the exploit was successful in 33.3\%, with the shellcode failing to execute in 66.7\% of tests. Tests where the exploit failed caused cFS to crash, resulting in an unintended DoS attack on the emulated satellite. 

\section{Discussion}
\label{sec:discussion}

\subsection{Scenario Assumptions and Realism}
There are three assumptions made in this work that allow the attack scenario to succeed. First, the that the attacker is able to reach the satellite while it is in-orbit. As mentioned in Section \ref{sec:related_work}, the accessibility of modern SDR equipment, and instances of functional GS locations being accessible and usable \cite{HackerBroadcast}, make satellite communication channels much more accessible than they historically have been. The second assumption is that a vulnerability is present in the custom cFS application. The vulnerability described in Section \ref{vuln-emu} is a typical RCE vulnerability that can easily be introduced by mismanaging memory in C programs. The third assumption is the use of Python on the satellite. While satellite software has previously involved minimal features running on custom firmware or an RTOS, modern CubeSats can involve more complex software stacks, with \cite{antunes2019developing} including both cFS and Python. The assumptions made in this paper are therefore realistic and based on real examples or existing work.   

\subsection{Exploit Reliability}
The reliability of the exploit was low with a success rate of 33.3\%. This was determined to be caused by a combination of the payload size and the stack usage at the time of exploitation varying between experiments. Improving reliability will require both minimising the shellcode and using an alternative exploitation method, such as Return Oriented Programming (ROP), which would have the added benefit of bypassing software security features that may be present on the system. 

\subsection{Defence}
The most suitable defence against this attack is to encrypt the data communication, preventing an attacker from sending their own messages to the target system. This was implemented a tested using hardware encryption supported by the radios used in the tested. This method, however, may not be suitable to real-world satellite deployments as encryption adds a layer of complexity to the data and software stack, Moreover, some systems may be required to be open and accessible to many GS and terrestrial systems.

\subsection{Future Work}
The current iteration of this work serves as a prototype for the presented exploit path and ransomware. Additional work will aim to refine the exploit, shellcode, and ransomware presented in this paper. A key goal of future work is to generalise the attack and ransomware for cFS as much as possible and reduce the reliance on Python for the malware. This will involve further investigation of cFS to identify ROP chains that could potentially embed the ransomware functionality within the shellcode. This refinement would also support the execution of this attack against systems running cFS on various other operating systems.

\section{Conclusion}
Satellite systems and NTNs play a pivotal role in the modern world, enabling global communications, navigation, weather monitoring, scientific research. This paper examined an exploit path that can allow ransomware to take control of an in-orbit satellite. This work presented a realistic scenario which left a satellite vulnerable to remote attackers. Using custom shellcode and novel ransomware, the attackers in the considered scenario could successfully exploit the vulnerability and prevent regular communication between the satellite and GS operator. The reliability of this attack was evaluated, showing that the ransomware was successfully delivered and executed in 33.3\% of experiments, with the remaining 66.7\% resulting in a DoS attack on the satellite. Future work will aim to refine the attack and ransomware to improve the exploit's reliability. 

%With the increase in popularity and power of SDRs, and their reduction in cost, the construction of rogue ground stations will become a threat to the security of satellites. For the longest time it was believed that once a vessel is launched into space, it would be out of the reach of attackers, while hiding the technology behind the spacecraft in the form of trade secrets in a "security-through-obscurity" fashion. In the threat of ransomware in space, this can become a serious disadvantage as once the operators lose contact with the satellite, it would become unrecoverable in space. 

%This work presented a concept for attacking a satellite, which was already launched into space by abusing a typical buffer overflow exploit. It further shows a concept of holding the vessel for ransom by targeting the communication module, ignoring commands from the legitimate operators and only releasing the vessel when a special key is received, which only the attacker has access to. For testing and development, a testbed was configured using a Raspberry Pi and Elegoo Uno R3 board, with radio communication. NASA's cFS was chosen as the flight software, while OpenC3 COSMOS was used for the legitimate ground software. The attacker interfaces with their rogue ground station using Python and uses radio communication to deliver the payload.

\label{sec:conclusion}

\bibliographystyle{IEEEtran}
\bibliography{main}

\end{document}